\documentclass[twocolumn]{emulateapj}

\usepackage{natbib}
\usepackage{epsfig}
\usepackage{amssymb,amsmath}
\usepackage{color}
\usepackage{xspace}

\newcommand{\msun}{{\rm M}_{\odot}}
\newcommand{\lsun}{{\rm L}_{\odot}}
\newcommand{\kms}{\, {\rm km\, s}^{-1}}

\newcommand{\kpc}{\, {\rm kpc}}

\newcommand{\der}{{\rm d}}

\newcommand{\mypm}[2]{^{+#1}_{-#2}}
\newcommand{\eg}{e.g.,}
\newcommand{\zl}{z_{\rm l}}
\newcommand{\zs}{z_{\rm s}}

\def\lname{HERMES~J105751.1+573027}
\def\lnameII{HLSW--01}
\def\kband{$K_{\rm p}$}

\begin{document}

\title{Modeling of the \lname~submillimeter source lensed by a dark matter dominated foreground group of galaxies.}
  
\shorttitle{Lens model of \lnameII}

\shortauthors{Gavazzi et~al.}

\author{R.~Gavazzi\altaffilmark{1},
A.~Cooray\altaffilmark{2,3},
A.~Conley\altaffilmark{4},
J.E.~Aguirre\altaffilmark{5},
A.~Amblard\altaffilmark{2},
R.~Auld\altaffilmark{6},
A.~Beelen\altaffilmark{7},
A.~Blain\altaffilmark{3},
R.~Blundell\altaffilmark{8},
J.~Bock\altaffilmark{3,9},
C.M.~Bradford\altaffilmark{3,9},
C.~Bridge\altaffilmark{3},
D.~Brisbin\altaffilmark{10},
D.~Burgarella\altaffilmark{11},
P.~Chanial\altaffilmark{12},
E.~Chapin\altaffilmark{13},
N.~Christopher\altaffilmark{14},
D.L.~Clements\altaffilmark{15},
P.~Cox\altaffilmark{16},
S.G.~Djorgovski\altaffilmark{3},
C.D.~Dowell\altaffilmark{3,9},
S.~Eales\altaffilmark{6},
L.~Earle\altaffilmark{17},
T.P.~Ellsworth-Bowers\altaffilmark{4},
D.~Farrah\altaffilmark{18},
A.~Franceschini\altaffilmark{19},
H.~Fu\altaffilmark{3},
J.~Glenn\altaffilmark{4},
E.A.~Gonz\'alez~Solares\altaffilmark{20},
M.~Griffin\altaffilmark{6},
M.A.~Gurwell\altaffilmark{8},
M.~Halpern\altaffilmark{13},
E.~Ibar\altaffilmark{21},
R.J.~Ivison\altaffilmark{21,22},
M.~Jarvis\altaffilmark{23},
J.~Kamenetzky\altaffilmark{17},
S.~Kim\altaffilmark{2},
M.~Krips\altaffilmark{16},
L.~Levenson\altaffilmark{3,9},
R.~Lupu\altaffilmark{5},
A.~Mahabal\altaffilmark{3},
P.D.~Maloney\altaffilmark{17},
C.~Maraston\altaffilmark{24},
L.~Marchetti\altaffilmark{19},
G.~Marsden\altaffilmark{13},
H.~Matsuhara\altaffilmark{25},
A.M.J.~Mortier\altaffilmark{15},
E.~Murphy\altaffilmark{3,26},
B.J.~Naylor\altaffilmark{9},
R.~Neri\altaffilmark{16},
H.T.~Nguyen\altaffilmark{9,3},
S.J.~Oliver\altaffilmark{18},
A.~Omont\altaffilmark{1},
M.J.~Page\altaffilmark{27},
A.~Papageorgiou\altaffilmark{6},
C.P.~Pearson\altaffilmark{28,29},
I.~P{\'e}rez-Fournon\altaffilmark{30,31},
M.~Pohlen\altaffilmark{6},
N.~Rangwala\altaffilmark{4},
J.I.~Rawlings\altaffilmark{27},
G.~Raymond\altaffilmark{6},
D.~Riechers\altaffilmark{3,32},
G.~Rodighiero\altaffilmark{19},
I.G.~Roseboom\altaffilmark{18},
M.~Rowan-Robinson\altaffilmark{15},
B.~Schulz\altaffilmark{3,26},
Douglas~Scott\altaffilmark{13},
K.S.~Scott\altaffilmark{5},
P.~Serra\altaffilmark{2},
N.~Seymour\altaffilmark{27},
D.L.~Shupe\altaffilmark{3,26},
A.J.~Smith\altaffilmark{18},
M.~Symeonidis\altaffilmark{27},
K.E.~Tugwell\altaffilmark{27},
M.~Vaccari\altaffilmark{19},
E.~Valiante\altaffilmark{13},
I.~Valtchanov\altaffilmark{33},
A.~Verma\altaffilmark{14},
J.D.~Vieira\altaffilmark{3},
L.~Vigroux\altaffilmark{1},
L.~Wang\altaffilmark{18},
J.~Wardlow\altaffilmark{2},
D.~Wiebe\altaffilmark{13},
G.~Wright\altaffilmark{21},
C.K.~Xu\altaffilmark{3,26},
G.~Zeimann\altaffilmark{34},
M.~Zemcov\altaffilmark{3,9},
J.~Zmuidzinas\altaffilmark{3,9}}
\altaffiltext{1}{Institut d'Astrophysique de Paris, UMR 7095, CNRS, UPMC Univ. Paris 06, 98bis boulevard Arago, F-75014 Paris, France}
\altaffiltext{2}{Dept. of Physics \& Astronomy, University of California, Irvine, CA 92697}
\altaffiltext{3}{California Institute of Technology, 1200 E. California Blvd., Pasadena, CA 91125}
\altaffiltext{4}{Dept. of Astrophysical and Planetary Sciences, CASA 389-UCB, University of Colorado, Boulder, CO 80309}
\altaffiltext{5}{Department of Physics and Astronomy, University of Pennsylvania, Philadelphia, PA 19104}
\altaffiltext{6}{Cardiff School of Physics and Astronomy, Cardiff University, Queens Buildings, The Parade, Cardiff CF24 3AA, UK}
\altaffiltext{7}{Institut d'Astrophysique Spatiale (IAS), b\^atiment 121, Universit\'e Paris-Sud 11 and CNRS (UMR 8617), 91405 Orsay, France}
\altaffiltext{8}{Harvard-Smithsonian Center for Astrophysics, 60 Garden Street, Cambridge, MA 02138}
\altaffiltext{9}{Jet Propulsion Laboratory, 4800 Oak Grove Drive, Pasadena, CA 91109}
\altaffiltext{10}{Space Science Building, Cornell University, Ithaca, NY, 14853-6801}
\altaffiltext{11}{Laboratoire d'Astrophysique de Marseille, OAMP, Universit\'e Aix-Marseille, CNRS, 38 rue Fr\'ed\'eric Joliot-Curie, 13388 Marseille cedex 13, France}
\altaffiltext{12}{Laboratoire AIM-Paris-Saclay, CEA/DSM/Irfu - CNRS - Universit\'e Paris Diderot, CEA-Saclay, pt courrier 131, F-91191 Gif-sur-Yvette, France}
\altaffiltext{13}{Department of Physics \& Astronomy, University of British Columbia, 6224 Agricultural Road, Vancouver, BC V6T~1Z1, Canada}
\altaffiltext{14}{Department of Astrophysics, Denys Wilkinson Building, University of Oxford, Keble Road, Oxford OX1 3RH, UK}
\altaffiltext{15}{Astrophysics Group, Imperial College London, Blackett Laboratory, Prince Consort Road, London SW7 2AZ, UK}
\altaffiltext{16}{Institut de RadioAstronomie Millim\'etrique, 300 Rue de la Piscine, Domaine Universitaire, 38406 Saint Martin d'H\`eres, France}
\altaffiltext{17}{Center for Astrophysics and Space Astronomy, University of Colorado, Boulder, CO 80309}
\altaffiltext{18}{Astronomy Centre, Dept. of Physics \& Astronomy, University of Sussex, Brighton BN1 9QH, UK}
\altaffiltext{19}{Dipartimento di Astronomia, Universit\`{a} di Padova, vicolo Osservatorio, 3, 35122 Padova, Italy}
\altaffiltext{20}{Institute of Astronomy, University of Cambridge, Madingley Road, Cambridge CB3 0HA, UK}
\altaffiltext{21}{UK Astronomy Technology Centre, Royal Observatory, Blackford Hill, Edinburgh EH9 3HJ, UK}
\altaffiltext{22}{Institute for Astronomy, University of Edinburgh, Royal Observatory, Blackford Hill, Edinburgh EH9 3HJ, UK}
\altaffiltext{23}{Centre for Astrophysics Research, University of Hertfordshire, College Lane, Hatfield, Hertfordshire AL10 9AB, UK}
\altaffiltext{24}{ Institute of Cosmology and Gravitation, University of Portsmouth, Dennis Sciama Building, Burnaby Road, Portsmouth PO1 3FX, UK}
\altaffiltext{25}{Institute for Space and Astronautical Science, Japan Aerospace and Exploration Agency, Sagamihara, Kana- gawa 229-8510, Japan}
\altaffiltext{26}{Infrared Processing and Analysis Center, MS 100-22, California Institute of Technology, JPL, Pasadena, CA 91125}
\altaffiltext{27}{Mullard Space Science Laboratory, University College London, Holmbury St. Mary, Dorking, Surrey RH5 6NT, UK}
\altaffiltext{28}{Space Science \& Technology Department, Rutherford Appleton Laboratory, Chilton, Didcot, Oxfordshire OX11 0QX, UK}
\altaffiltext{29}{Institute for Space Imaging Science, University of Lethbridge, Lethbridge, Alberta, T1K 3M4, Canada}
\altaffiltext{30}{Instituto de Astrof{\'\i}sica de Canarias (IAC), E-38200 La Laguna, Tenerife, Spain}
\altaffiltext{31}{Departamento de Astrof{\'\i}sica, Universidad de La Laguna (ULL), E-38205 La Laguna, Tenerife, Spain}
\altaffiltext{32}{Hubble Fellow}
\altaffiltext{33}{Herschel Science Centre, European Space Astronomy Centre, Villanueva de la Ca\~nada, 28691 Madrid, Spain}
\altaffiltext{34}{University of California, 1 Shields Ave, Davis, CA 95616}

\email{gavazzi@iap.fr}

\begin{abstract} 
We present the results of a gravitational lensing analysis of the
bright $\zs=2.957$ sub-millimeter galaxy (SMG), \lname\ found in the
{\it Herschel}/SPIRE Science Demonstration Phase data from the
Herschel Multi-tiered Extragalactic Survey (HerMES) project.  The high
resolution imaging available in optical and Near-IR channels, along
with CO emission obtained with the Plateau de Bure Interferometer,
allow us to precisely estimate the intrinsic source extension and
hence estimate the total lensing magnification to be $\mu=10.9\pm
0.7$. We measure the half-light radius $R_{\rm eff}$ of the source in
the rest-frame Near-UV and $V$ bands that characterize the unobscured
light coming from stars and find $R_{\rm eff,*}= [2.0 \pm 0.1] \kpc$,
in good agreement with recent studies on the Sub-Millimeter Galaxies
population.  This lens model is also used to estimate the size of the
gas distribution ($R_{\rm eff,gas}= [1.1\pm0.5]\kpc$) by mapping back
in the source plane the CO ($J=5\rightarrow4$) transition line
emission.  The lens modeling yields a relatively large Einstein radius
$R_{\rm Ein}= 4\farcs10 \pm 0\farcs02$, corresponding to a deflector
velocity dispersion of [$483\pm 16] \,\kms$.  This shows that
\lname~is lensed by a {\it galaxy group-size} dark matter halo at
redshift $\zl\sim 0.6$. The projected dark matter contribution largely
dominates the mass budget within the Einstein radius with $f_{\rm
  dm}(<R_{\rm Ein})\sim 80\%$.  This fraction reduces to $f_{\rm
  dm}(<R_{\rm eff,G1}\simeq 4.5\kpc )\sim 47\%$ within the effective
radius of the main deflecting galaxy of stellar mass $M_{\rm
  *,G1}=[8.5\pm 1.6] \times 10^{11}\msun$.  At this smaller scale the
dark matter fraction is consistent with results already found for
massive lensing ellipticals at $z\sim0.2$
from the SLACS survey.
\end{abstract}


\section{Introduction}\label{sec:intro}
The current generation of wide field surveys at sub-millimeter and
millimeter wavelengths is now providing us with large numbers of high
redshift galaxies containing large amounts of dust heated by intense
star formation or AGN activity
\citep[\eg][]{hughes98,barger98,blain02,chapman05,coppin06,austermann10}.
This population of sub-millimeter galaxies (SMGs) is easily detectable
in the redshift range $1\lesssim z\lesssim 5$ thanks to a strong
negative k-correction when observed at wavelengths $\lambda\gtrsim
500\,\mu$m. This property, along with a sharp fall-off at the bright
luminosity end of their luminosity function, makes bright SMGs good
candidates for being strongly gravitationally lensed
\citep[\eg][]{blain96,negrello07,cooray10,negrello10,vieira10}.
Efficient identification of lensed SMGs is only now becoming possible
thanks to surveys like the Herschel Multi-tiered Extragalactic Survey
(HerMES) or the Herschel - Astrophysical Terahertz Large Area
Survey(H-ATLAS) conducted with the {\it Herschel} satellite
\citep[\eg][]{oliver10,griffin10,negrello10} and from the ground with
the South Pole Telescope \citep{vieira10}.

The interest of building large samples of lensed SMGs for getting
better insights on the properties of these otherwise very faint
objects is clear, and recent results are already shedding some light
on the spatial distribution of gas, dust and stars in these SMGs
\citep[\eg][]{swinbank10b}. However the redshift distribution of SMGs,
which peaks in the range $2-2.5$ \citep{chapman05,wardlow10}, is also
well suited to probe the mass distribution of high redshift
deflectors, typically in the range $0.3<\zl<1.5$, which complements
local studies like the Sloan Lens ACS Survey (SLACS) which are limited
to $z\le 0.4$ \citep{bolton08a,auger10}.

In this paper, we present the modeling of the gravitationally lensed
SMG \lname, also referred to as \lnameII, discovered in {\it
  Herschel}/SPIRE observations during the Science Demonstration Phase
by the HerMES program. Its J2000 coordinates are RA=10:57:51.0,
Dec=+57:30:25.8 with a lens photometric redshift of $\zl=0.60\pm0.04$
\citep{oyaizu08} and a source redshift of $\zs=2.9575\pm0.0001$
(Riechers et~al.\ 2011; Scott et~al.\ 2011; hereafter R11 and S11).
The goal of this lens modeling work is two-fold. First we want to
recover the intrinsic light distribution of the source, while
optimally taking advantage of the magnifying power of the
deflector. Detailed investigations on the lensed source are developed
in associated papers (Conley et al. 2011 [hereafter C11], R11, S11).
And second we want to probe the mass content of the foreground object
which, given the large image separation of the multiple images, might
be very massive.

The paper is thus organized as follows. In \S\ref{sec:optical} we
present the lens modeling techniques and the optical and Near-IR data
we shall use along with the main results. The best fit lens model is
then used in \S\ref{sec:submm} to reconstruct the source
CO($J=5\rightarrow4$) light distribution at 576 GHz observed with the
Plateau de Bure Interferometer (PdBI). In \S\ref{sec:disc}, we
interpret the lens model results to measure the balance of dark and
luminous matter in the inner $30 \kpc$ of the deflector. We conclude
in \S\ref{sec:conc}. Throughout, we assume a concordance cosmology
with matter and dark energy density $\Omega_{\rm m}=0.3$,
$\Omega_{\Lambda}=0.7$, and Hubble constant $H_0$=70 km
s$^{-1}$Mpc$^{-1}$. Magnitudes are expressed in the AB system.

\section{Lens modeling of optical data}\label{sec:optical}
\subsection{Observations}\label{sec:data}
For accurate lens modeling, we use the best spatial resolution images
of sufficient signal-to-noise that are currently available for
\lnameII.  A 1400 second image of the system was taken using Laser
Guide Star Adaptive Optics in the \kband~band with the NIRC2
instrument mounted on the Keck~II
telescope\footnote[1]{\url{http://www2.keck.hawaii.edu/realpublic/inst/nirc2/index.html}}.
The observing conditions allowed us to achieve a typical $0\farcs2$
FWHM point spread function.  The FWHM is well sampled with a plate
scale of $0\farcs02$.  We additionally use a 1 hour Subaru SuprimeCam
$i$ band observation with $\sim0\farcs74$ FWHM seeing and $0\farcs202$
pixel size.

\begin{figure}
\centering\includegraphics[width=0.95\linewidth]{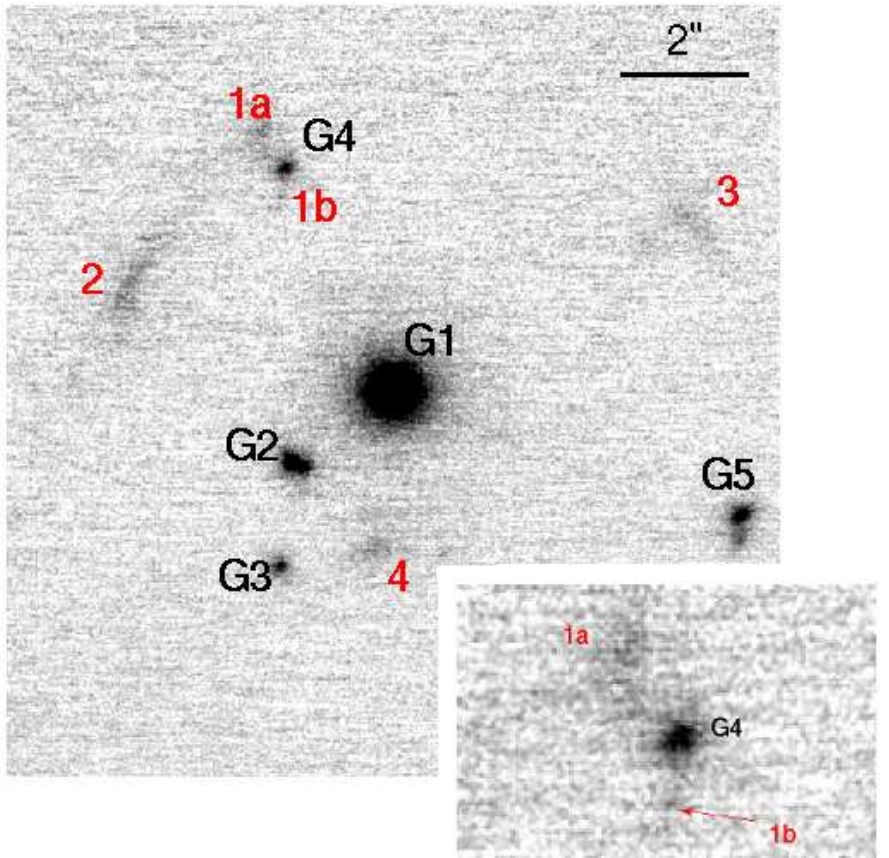}\\
\caption{\label{fig:general} \kband~band adaptive optics-corrected
  overview of \lnameII~with labeled multiple images (red) and
  foreground galaxies (black). North is up and East is left. The
  bottom right inset zooms into the image 1 that appears to be
  perturbed by the foreground galaxy G4 and split into two pieces.}
\end{figure}

A \kband~image of the central region of the system is shown in
Fig.~\ref{fig:general}.  The lensing configuration can easily be
described despite the relatively low surface brightness of the
multiply-imaged features of \lnameII. The configuration of multiple
images lifts any ambiguities as we clearly see four images of similar
surface brightness forming a, so-called, {\it fold} configuration with
images 1 and 2 presumably merging through the critical line and images
3 and 4 being lower magnification {\it conjugate} images \citep[see
  \eg][for a description of catastrophe theory in the context of
  gravitational lensing]{SEF92}.  A closer look at the \kband~band
image shows that image 1 is perturbed by a small galaxy, G4, which
seems to be massive enough to split image 1 into two pieces (1a and
1b) on both sides of G4. Galaxies G2 and G3 may also act as potential
perturbers on the innermost multiple image, 4. To a lesser extent, G5
might also be considered as a perturbing galaxy. G1 is the central
galaxy of the massive deflecting structure, which presumably should be
a group of galaxies, given the large $\Delta \theta \simeq 8\arcsec$
image separation. Having no redshift information at all for these
perturbers, the simplest assumption is to consider them to be at the
same redshift as G1, keeping in mind that changes from this hypothesis
are of second order in the lens modeling. Indeed, the total
magnification of \lnameII~is not changed, while the absolute mass
calibration of G2...G5 depends on the assumption that they are at the
same redshift as G1. However, measuring the mass of perturbers is not
the main motivation of the modeling.

\subsection{Method}\label{sec:method}
The lens modeling builds on the dedicated code {\tt sl\_fit}
previously used for galaxy-scale strong lenses
\citep[\eg][]{gavazzi07b,gavazzi08,ruff10}. It fits model parameters
of simple analytic lensing potentials. {\tt sl\_fit} can be run in
three different regimes of increasing computational cost. The first
mode makes use of the coordinates of image plane points and minimizes
the distance to their parent source plane locations in a way similar
to {\tt gravlens} \citep{keeton01soft1} or {\tt lenstool}
\citep{kneib93these,jullo07}. The second mode uses the full surface
brightness distribution and attempts to account for it with one or
more simple analytic light components that we take to have a unique
Gaussian radial profile with elliptical shape \citep[see \eg][for
  similar techniques]{marshall07,bolton08a}. Finally the third mode
implements a pixellized linear reconstruction of the source plane
light distribution while fitting for the non-linear potential
parameters \citep{warren03,treu04,suyu06}; we did not consider the
latter mode here as its computational cost is prohibitive for the
large images and complex gravitational potential of this system.

The lensing potential is assumed to be made of a cored isothermal
ellipsoid, centered on the main deflector galaxy G1, and that is
supposed to capture the lensing contribution of the dark matter halo
as well as the stellar component of G1. Given the absence of a radial
arc or central demagnified images \citep[see \eg][]{gavazzi03}, the
details of the assumed potential in the innermost parts ($r\lesssim
2\arcsec $) of the lens should not be important. The peak of G1's
light distribution is assumed to be the center of this potential
component. The convergence profile of the central mass component is
given by
\begin{equation}\label{eq:main}
\kappa_{\rm cent}(\vec{\xi}) = \frac{b_{\rm cent}}{2} \frac{1}{\sqrt{ \xi^2+r_c^2}}\,,
\end{equation}
where the scaling parameter $b_{\rm cent}$ is related to the velocity
dispersion of the deflector through $b/1\arcsec = (\sigma_v/186.21
\,\kms)^2 D_{\rm ls}/D_{\rm s}$, the core radius is $r_c$,
$\xi^2=x^2+y^2/q^2$ is the radial coordinate that accounts for the
ellipsoidal symmetry of the isodensity contours and $q$ is the
minor-to-major axis ratio. The orientation of the major axis,
$\theta_{\rm cent}$, relative to the $x$-axis is allowed to vary,
although this is not explicit in the definition of $\vec{\xi}$. As
will be seen below, lens modeling of extended images can yield formal
errors on $b$ of order one percent, and therefore similar errors on
$\sigma_v$. However, here we propagate uncertainties in $D_{\rm
  ls}/D_{\rm s}$ due to the relatively poorly known lens redshift
($\zl=0.60\pm0.04$). This results in a dominant additional 3\% error
that we add in quadrature to model uncertainties on $b$, keeping in
mind that a spectroscopic measurement of $\zl$ would readily reduce
lens modeling errors on $\sigma_v$ to the percent level.

We carefully account for the perturbing galaxies G2, G3, G4 and the
more distant G5, and model them as truncated isothermal ellipsoids,
each with a free velocity dispersion and a constant cutoff radius that
we fix to a value of $r_{\rm t}=1\farcs5$ corresponding to $10 \kpc$
\citep[see \eg][]{limousin07,limousin09a,suyu10}.  For each of these
galaxies the convergence profile takes the form:
\begin{equation}
\kappa_{\rm pert}(\vec{r}-\vec{r_{\rm G,i}}) = \frac{b_{\rm pert,i}}{2} \left( \frac{1}{\xi}- \frac{1}{\sqrt{\xi^2+r_t^2}}\right)\;,
\end{equation}
so that the total mass of a perturber is $M_{\rm pert,i}= \pi
\Sigma_{\rm crit} b_{\rm pert,i} r_t$ where $\Sigma_{\rm crit}$ is the
critical surface density, which for this system has a value
$\Sigma_{\rm crit}\simeq [1.80 \pm 0.05] \times 10^{9}
\,\msun\,\kpc^{-2} \simeq [8.05 \pm 0.24] \times 10^{10}\,\msun\,{\rm
  arcsec}^{-2}$.  The 3\% quoted errors are dominated by the
uncertainty in the photometric redshift of the lens, but this is still
negligible compared to the uncertainty in the velocity dispersion of
the main component.  As an approximation, we match the ellipticity and
orientation of these perturbing systems to that of the stars they
host, although we note that recent analyses have shown that departures
from this simple assumption might occur and is sometimes observed with
more favorable lensing configurations and deep space-based optical
data\citep{suyu10}.

To summarize, the lensing potential is described by 4 parameters for
the central component: $b_{\rm cent}$, $r_c$, $q_{\rm cent}$,
$\theta_{\rm cent}$ and we also allow the 4 scaling parameters $b_{\rm
  pert,i=2\ldots5}$ of the perturbing potentials to vary.

\subsection{Preliminary analysis:  conjugate points}\label{sseq:conj}
We first attempted to fit the model parameters of the main central
component assuming that the source is point-like (first mode of {\tt
  sl\_fit}) and the perturbers have a negligible effect.  By ignoring
the effect of G4 we had to neglect the additionally produced image 1b
and just considered the brightest pixel of each of the images 1a, 2, 3
and 4 as input constraints on the potential. We performed a fit to the
parameters $q_{\rm cent}$, $\theta_{\rm cent}$, $b_{\rm cent}$ and
$r_c$. We used these results to further understand the source
structure (fainter spots) seen in image 2 and how it is cast onto
other images 1a, 3 and 4 respectively, and, hence increase the number
of constraints \citep[see \eg][for an illustration of the
  process]{gavazzi03}. This also allowed us to determine which part of
the source should be inside the extra caustic caused by the perturber
G4 and thus seen as image 1b. Altogether we identified 4
multiply-imaged knots.

We ran a Monte-Carlo Markov Chain sampler of the posterior
distribution of the 8 model parameters related to the gravitational
potential. At this stage the most important results\footnote[2]{For a
  given parameter we quote the median and 68\% confidence level errors
  as given by the 16th and 84th percentiles of the marginalized
  distribution.} are that we find quite an elongated mass distribution 
with an axis ratio $q_{\rm cent}=0.41\pm 0.03$ and orientation
$\theta_{\rm cent}= 13\fdg 8  \pm 0\fdg 5$ (North to East counterclockwise).
The circularized Einstein radius is $R_{\rm Ein}= 4\farcs02\pm0\farcs05$,
corresponding to a velocity dispersion of $\sigma_v = [473 \pm 15] \,\kms$,
which suggests that the deflector is a massive group of galaxies.
The modeling seems to require a finite core radius $r_c = [2.0 \pm 1.0] \kpc$,
but it is difficult to interpret this as a constraint on the dark matter
distribution since the central component of Eq.~(\ref{eq:main}) captures both
the contribution of stars and dark matter.
The mass of perturber G4 is relatively well constrained:
  $M_{G4} = [25 \pm 5] \times 10^{10} \msun$, whereas
  $M_{G2} = [13 \pm 9] \times 10^{10}\msun$,
  $M_{G3} = [64 \pm 6]\times 10^{10} \msun$ and 
  $M_{G5} = [20 \pm 12] \times 10^{10} \msun$.
G5 has a nearly negligible effect on the lensing configuration.

This best fit model predicts local magnifications\footnote[3]{Negative
 values indicate a change of image parity with respect to that of the
 source.} of $-$1.87, 4.37, 2.57 and $-$1.86 for images 1b, 2, 3 and 4,
respectively, which means that the total magnification should be
of order 10 for a point-like source. However the above treatment of
point-like lensing observables, although it allowed us to quickly explore
the space of lens potential parameters, does not take full advantage
of the spatially resolved surface brightness distribution. 
It thus cannot give us a clear idea of the intrinsic source light distribution,
on which the actual magnification factor depends.

\begin{figure}
\centering\includegraphics[width=0.99\linewidth]{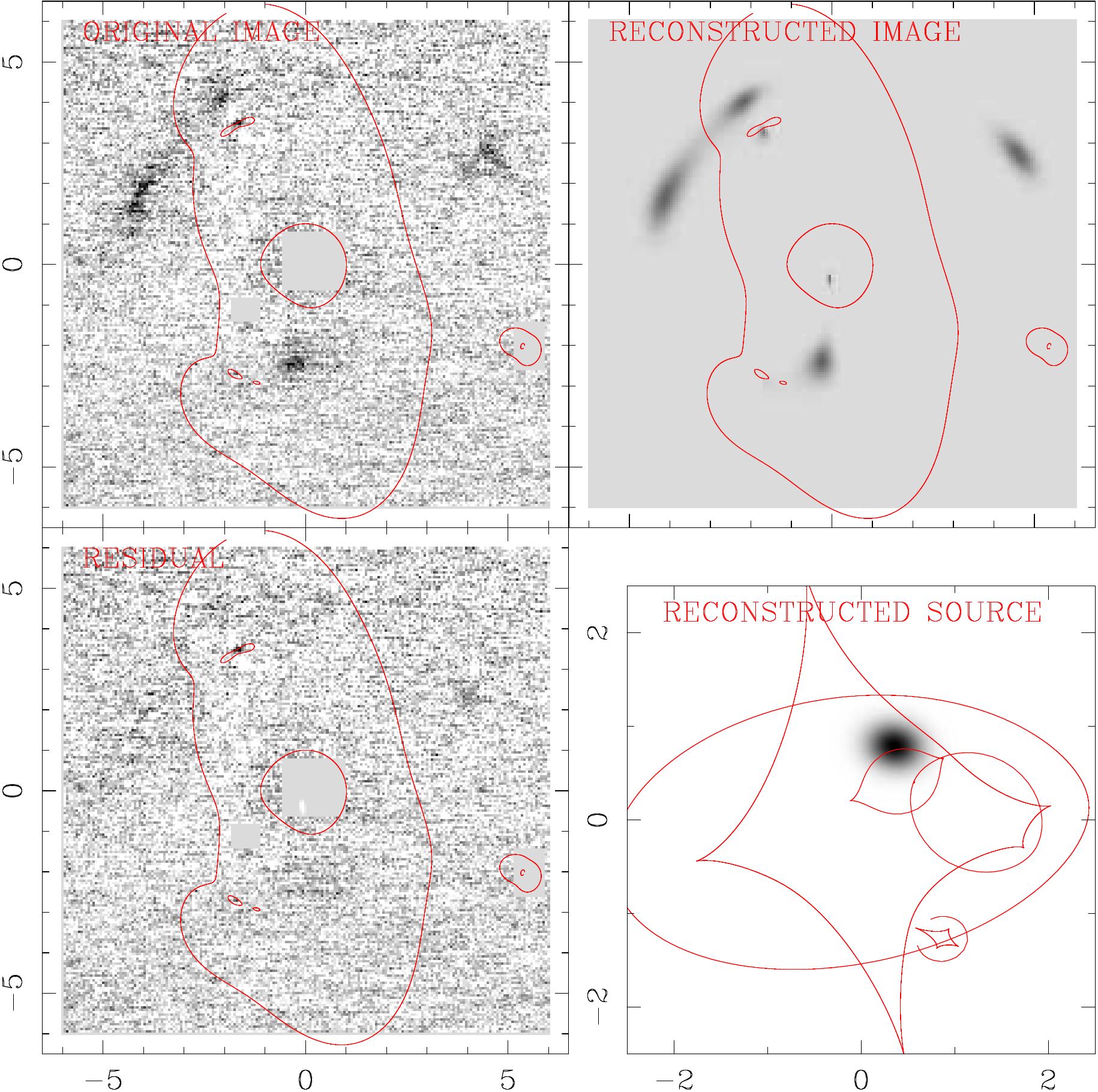}\\
\caption{\label{fig:KK} Results of the surface brightness fit of the
  Keck LGSAO \kband~band imaging. {\it Top left panel:} Input image
  with the foreground deflectors subtracted off (with their core
  completely masked out in some cases) using {\tt galfit}. {\it Top
    right panel:} Image plane model predictions. {\it Bottom left
    panel:} Image plane residuals (data-model). {\it Bottom right
    panel:} Source plane model predictions. In this latter panel, the
  caustic lines are shown in red, whereas the other panels show the
  image plane critical lines.  Scales are given in arc seconds.}
\end{figure}

\subsection{Keck LGSAO \kband~band light distribution}\label{sseq:Kext}
We therefore placed an extended source with an analytic light profile
made of an elliptical Gaussian with free central position ($x_{\rm
  s}$, $y_{\rm s}$), axis ratio ($q_{\rm s}$), position angle
$\theta_{\rm s}$, flux $F_{\rm s}$ and half-light radius $R_{\rm
  eff,s}$.  We optimized these parameters along with the potential
parameters already introduced in the previous exploration phase using
the surface-brightness mode of {\tt sl\_fit}. Fig.~\ref{fig:KK} shows
the resulting fit in the image plane. The corresponding reconstructed
source is shown in the bottom right panel. The best fit is achieved
with an extended source of \kband~band magnitude $22.12\pm 0.08$ and
half-light radius $R_{\rm eff,s}= [1.88 \pm 0.09] \kpc$.

\begin{figure}
\centering\includegraphics[width=0.99\linewidth]{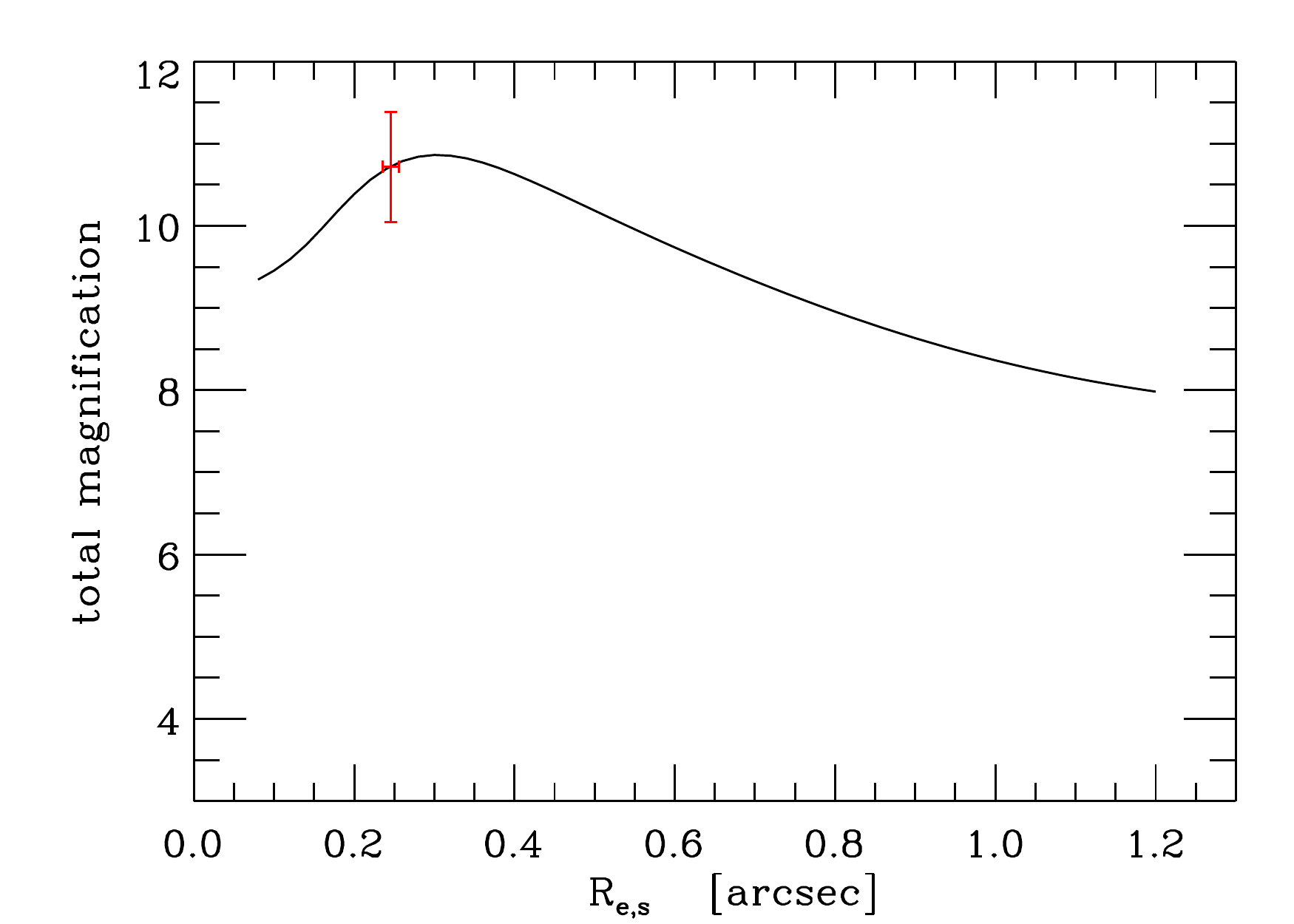}\\
\caption{\label{fig:magRe} Change of the total magnification as a
  function of the characteristic source size $R_{\rm eff,s}$. The error
  bar shows the statistical error inferred from fitting the \kband~band
  Keck LGSAO image (corresponding to a rest frame V band
  observation). The drop at low values of $R_{\rm eff,s}$ is
  artificially due to sampling limitations but does not affect the analysis.}
\end{figure}

 
We estimate the net magnification experienced by the source through a
direct numerical sum of pixel values in the image and the source
planes and find the magnification to be $\mu=10.86\pm 0.68$ after
marginalizing over source and potential parameters.  We show in
Fig.~\ref{fig:magRe} the change of the magnification as a function of
the source size $R_{\rm eff,s}$. This suggests that, provided the size
stays within a factor of $\sim 2$ for one band to another, the
systematic change in magnification remains within the current
statistical errors. We can also see that magnification is shown to be
$\lesssim 11.5$ whatever the source size and this can readily be cast
as a robust lower bound on the intrinsic source flux for a given total
observed flux.

Compared to the previous analysis using the conjugation of bright
knots, the fit of an extended source yields little change in the
recovered lens potential. The key features of the lens potential are
\begin{itemize}
\item The axis ratio of the mass distribution is $q_{\rm cent}=0.45\pm 0.02$,
  with orientation $\theta_{\rm cent}= 11\fdg9 \pm 0\fdg4 $.
\item The (circularized) Einstein radius is
  $R_{\rm Ein}= 4\farcs10\pm 0\farcs02 $, which corresponds to a velocity
  dispersion of $ \sigma_v = [483\pm 16 ]\,\kms$, in agreement with the
  preliminary study of \S\ref{sseq:conj}. The mass content of this deflecting component
  is further discussed in \S\ref{sec:disc}
\item The core radius $r_c= [2.7 \pm 0.8]\,\kpc$ is found to be slightly larger.
  It is mostly constrained by the size and shape of the innermost image 4.
\item There is little change in the mass of perturbers, for which we find:
  $M_{G4} = [26.8 \pm 1.8] \times 10^{10} \msun$,
  $M_{G2} = [3.2 \pm 2.4] \times 10^{10}\msun$,
  $M_{G3} = [49.4 \pm 5.2] \times 10^{10} \msun$ and 
  $M_{G5} = [12.7 \pm 7.0] \times 10^{10} \msun$.
\end{itemize}

\subsection{Seeing-limited Subaru $i$ band light distribution}\label{sseq:iext}
We now take advantage of the somewhat deeper seeing limited Subaru $i$
band image to investigate the ability of the \kband\ model to account
for observations at shorter wavelengths. This image corresponds to
rest frame $\sim$ 2000\AA~NUV emission that we expect to be lumpier
and severely obscured by dust, and thus not to have the same extent as
the rest frame V band probed by the \kband~band image.

\begin{figure}
\centering\includegraphics[width=0.99\linewidth]{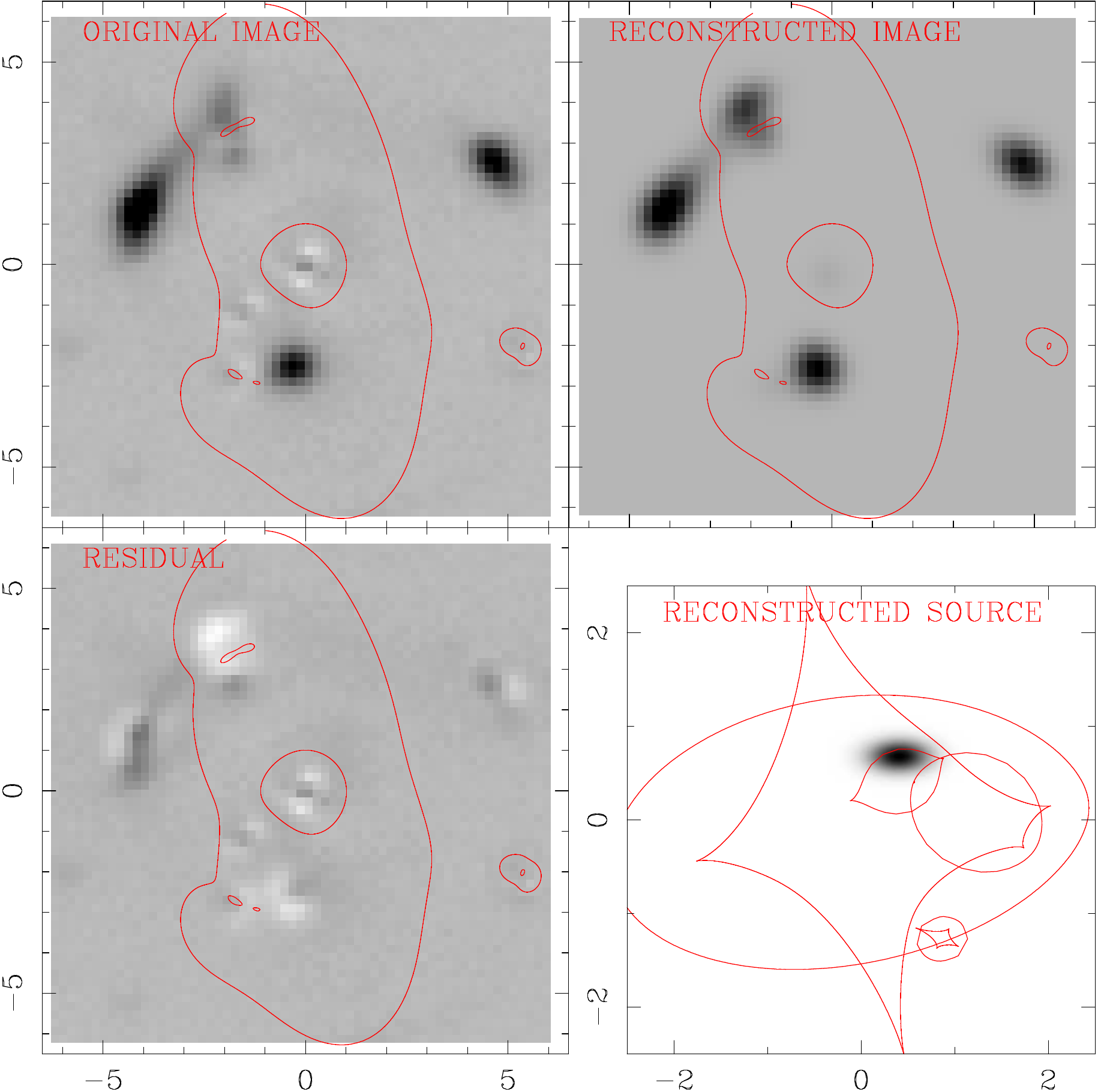}\\
\caption{\label{fig:subaru_i} Subaru $i$ band modeling results. {\it
    Top left panel:} input image with the foreground deflectors
  subtracted off. {\it Top right panel:} image plane model
  predictions. {\it Bottom left panel:} Image plane residuals
  (data-model). {\it Bottom right panel:} source plane model
  predictions.}
\end{figure}

We therefore repeated the previous analysis on the Subaru $i$ image,
but we considered the best fit potential values above and only
attempted to fit for the $i$ band light profile parameters of the
source: central position ($x_{\rm s}$, $y_{\rm s}$), axis ratio
($q_{\rm s}$), position angle $\theta_{\rm s}$, flux $F_{\rm s}$ and
half-light radius $R_{\rm eff,s}$.  Fig.~\ref{fig:subaru_i} shows the
result of the surface brightness distribution fit in the image
plane. The corresponding reconstructed source is shown in the bottom
right panel. The source is found to have a similar appearance in $i$
and \kband~bands with a half-light radius $R_{\rm eff,s}= [2.00 \pm
  0.01] \kpc$ and an $i$ band AB magnitude of $22.72 \pm 0.01$.  These
small statistical errors should be treated with caution, as
\kband~but the residuals are worse than the for \kband~case, which
might indicate a more complex intrinsic NUV light distribution. These
departures from a simple Gaussian elliptical profile could not be
observed in \kband~because of the substantially lower signal-to-noise
ratio.

This latter analysis demonstrates that lens model parameters found by
fitting the light distribution in the redder \kband~filter give
satisfying results for the fit of the light distribution in a
different band, $i$.

\section{CO(5-4) line distribution}\label{sec:submm}
Given the above success at accounting for the lensed visible/NIR light
distribution of \lnameII, we extend our analysis to sub-mm wavelengths
using spatially resolved PdBI observations of the
CO($J=5\rightarrow4$) transition line at 576 GHz (see R11 for
details).

Since our lens modeling code was originally designed to model
optical/NIR images, we could not fit these observations directly in
the uv-plane, which would allow a more careful handling of the beam
convolution and limit the effect of noise correlations. In order to
circumvent these difficulties we used a reduced image that was CLEANed
with a synthesized beam of $4\farcs8\times 2\farcs7$ FWHM, with a
major axis oriented $+51\fdg6$ East of North. The measured noise rms
is $1.2 \,{\rm mJy}\,{\rm beam}^{-1}$.

\begin{figure}
\centering\includegraphics[width=0.99\linewidth]{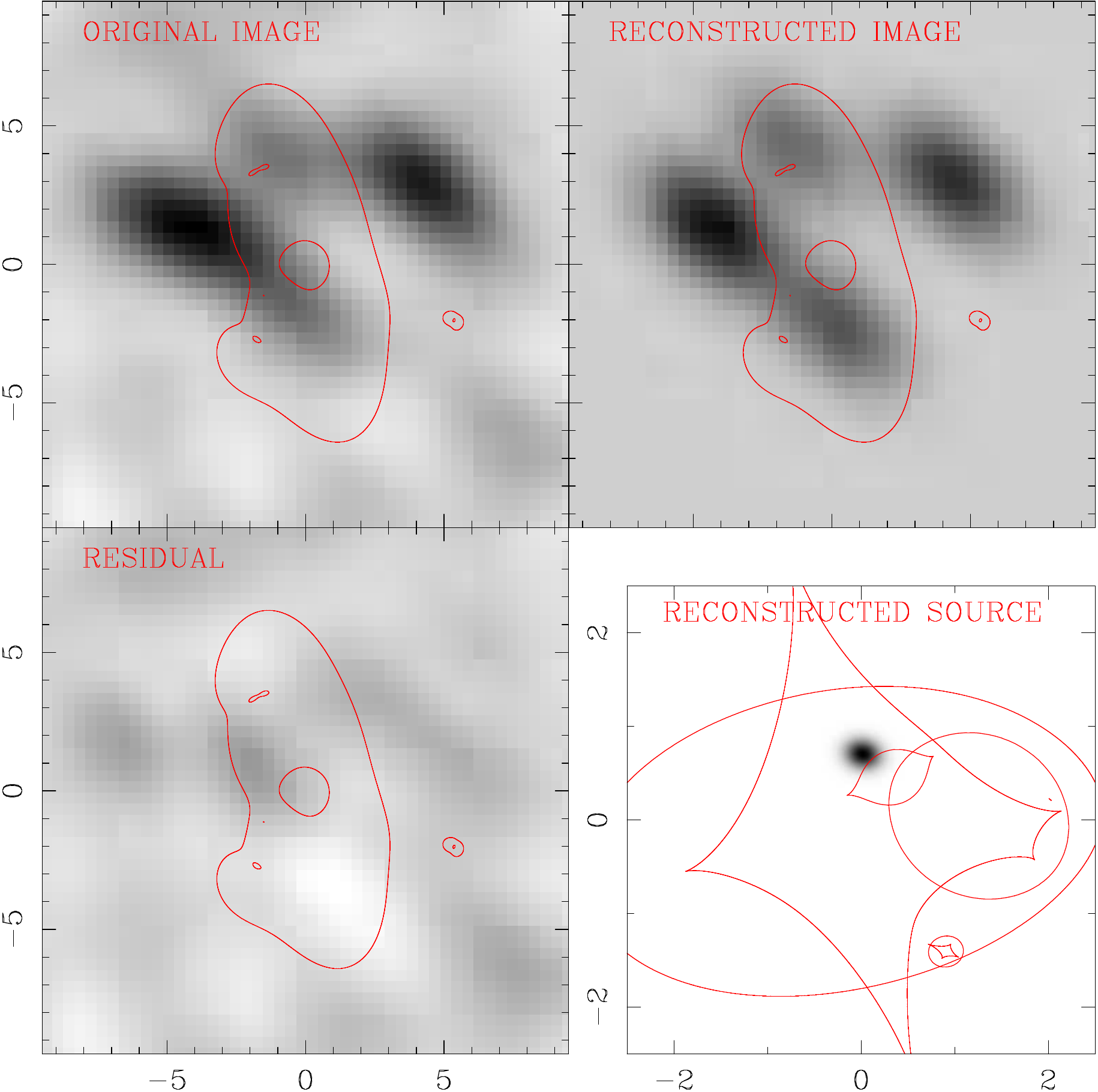}\\
\caption{\label{fig:PdBIflux} Plateau de Bure Interferometer
  CO($J=5\rightarrow4$) line flux density maps modeling results. {\it
    Top left panel:} input image with the foreground deflectors
  subtracted off. {\it Top right panel:} image plane model
  predictions. {\it Bottom left panel:} Image plane residuals
  (data-model). {\it Bottom right panel:} source plane model
  predictions. Note the slightly larger scale compared to
  Figs.~\ref{fig:KK} and \ref{fig:subaru_i}.}
\end{figure}

We did not attempt to fit for lens potential model parameters, as they
were better determined with visible/NIR data, but we fit for source
position integrated intrinsic flux and shape. Fig.~\ref{fig:PdBIflux}
shows the result of the CO($J=5\rightarrow4$) emission map fit in the
image plane.  The corresponding reconstructed source is shown in the
bottom right panel.  We can see that the modeling yields very small
residuals beyond the secondary lobes that cannot be captured by our
direct space modeling strategy.  Our main inferences for the source
parameters are
\begin{itemize}
\item We find the source half flux radius to be $R_{\rm eff,s}= [1.13\pm0.53] \kpc$,
  slightly smaller than the extent we found for the rest-frame V and NUV bands fits.
\item We measure a small offset between the center of the gas
  distribution as probed by the CO($J=5\rightarrow4$) transition and
  the stars that emit at visible/NIR wavelengths. The typical offset
  is $\Delta {\rm RA}=-0\farcs29\pm0\farcs07$ and $\Delta {\rm
    Dec}=-0\farcs10\pm 0\farcs05$ (corresponding to about $2.4\kpc$).
  This is suggestive that rest frame UV and visible light is severely
  obscured by the large dust content presumably associated with the
  gas revealed by the CO($J=5\rightarrow4$) line (R11). The presence
  of dust is confirmed by the Far IR emission (C11).  We stress
  however that further investigation of this is left for future work,
  including higher resolution mm or radio data.
\item The intrinsic source luminosity of the CO($J=5\rightarrow4$) line is
  $L^\prime_{CO(5-4)}=[3.76 \pm 0.44]\,\times 10^{10}$ K km s$^{-1}$ pc$^2$
  including the equally important instrumental and modeling error terms.
\end{itemize}

The good signal-to-noise ratio of the PdBI data suggests that one
could use the kinematical information to try and reconstruct the
intrinsic velocity field as traced by the CO($J=5\rightarrow4$)
transition line. However, because of insufficient spatial resolution
and perhaps non trivial patterns in the velocity field, we were not
able to explain the marginally resolved velocity field presented in
Fig.~5 of R11.

\section{Mass content of the deflector}\label{sec:disc}
We turn now to an analysis of our findings regarding the lens
potential and compare the inferred mass distribution with the
properties of the main deflector.

We used {\tt galfit} \citep{peng02} to measure the \kband~light
properties of the main galaxy G1 and found that the surface brightness
profile is well fit by a S\'ersic profile \citep{sersic68} of index
$n=3.06 \pm 0.10$, effective radius $R_{\rm eff,G1}= [4.5 \pm 0.1]
\kpc$, total \kband~magnitude\footnote[4]{Photometry is quite
  uncertain with adaptive optics imaging, and calibration was made
  difficult due to uncertainties in conversion between NIRC2
  \kband~filter and the reference 2MASS $K_{\rm s}$ photometry.}
$17.41 \pm 0.10$ and a magnitude $18.37\pm 0.04$ in the $i$
band. Assuming a spectral energy distribution typical of an elliptical
galaxy \citep{coleman80}, we can transform this observed magnitude
into a rest-frame V band luminosity $L_V = [3.9 \pm 0.6] \times
10^{11} \lsun $, including errors on photometry (4\%), photometric
redshift (16\%) and uncertainties in filter conversions (13\%).
Taking into account the luminosity decline due to the passive
evolution of its old stellar populations \citep{treu01},
\begin{equation}
  \frac{\der\;}{\der z} \log \frac{M_*}{L_V}   \simeq -0.40 \pm 0.05\;,
\end{equation}
we can predict the evolution-corrected luminosity. Applying the same
stellar {\it mass-to-light} ratio as found by \citet{gavazzi07b} for
massive early-type galaxies in the SLACS survey we get a total stellar
mass $M_{\rm *,G1}= 8.5\pm 1.6 \times 10^{11} \msun$ for the central
galaxy. This large stellar mass, along with the large characteristic
velocity dispersion inferred from lens modeling $\sigma_v = [483\pm
  16]\,\kms$, suggests that the deflector is likely a massive group of
galaxies with G1 being its dominating central galaxy\footnote[5]{We
  could not find any concentration of galaxies within several arcmin
  probed by the Subaru image, further indicating that G1 is the main
  mass component.}.

It is tempting to compare these results with the SLACS findings for
massive Early-Type Galaxies at redshift $z\sim0.2$ and with median
stellar mass $\sim 2-4 \times 10^{11}\msun$ \citep[depending on the
  choice of the stellar Initial Mass Function][]{auger10}.  We find
that our lens model requires a finite core radius $r_c=[2.72\pm 0.84]
\kpc$. This is substantially different from lensing findings in $z\sim
0.2$ elliptical galaxies of slightly lower mass
\citep[\eg][]{bolton08a,koopmans09} in which a Singular Isothermal
Elliptical profile provides a good description of the total lensing
potential.  In addition the lensing-inferred velocity dispersion is
unlikely to reflect the stellar velocity dispersion of G1 that would
be measured by optical/NIR spectroscopy. Indeed, the Fundamental Plane
(FP) of early-type galaxies \citep{djorgovski87,faber87} predicts that
the velocity dispersion of G1 should be $[400 \pm 30]\,\kms$ with the
latest normalization of \citet{auger10}. However the finite core
radius requires a numerical integration of the Jeans equation to
relate our total density profile inferred from lensing and a true
velocity dispersion $\sigma_{\rm ap}$ within a fiducial aperture of
radius $R_{\rm ap}=1\farcs5 \simeq 10 \kpc$.  Following the
prescriptions of \citet{mamon05b} for fast numerical integrations of
the Jeans equation, assuming an isotropic velocity tensor, our model
predicts a value $\sigma_{\rm ap} = [380 \pm 5] \,\kms$, in good
agreement with the FP value.

A direct comparison with SLACS is complicated by the very different
values of Einstein radii.  For SLACS the ratio $R_{\rm Ein}/R_{\rm
  eff}\sim 0.5$ whereas we observe $R_{\rm Ein}\sim 27 \kpc \sim
6\times R_{\rm eff}$. Therefore, it is not surprising that the total
mass within $R_{\rm Ein}$ is a factor of $\sim 5$ greater than the
stellar mass\footnote[6]{This corresponds to a total {\it
    mass-to-light} ratio of $M_{\rm tot}/L_V = 18.0\mypm{3.6}{2.7}
  \left(M/L_V\right)_\sun$, again, correcting for luminosity
  evolution}. The deflector is thus a group of galaxies that is more
dark matter dominated than most SLACS massive early-type galaxies. We
note that the cosmological critical density for lensing $\Sigma_{\rm
  crit} = c^2/(4\pi G) D_{\rm s} / (D_{\rm ls} D_{\rm os})$ is more
favorable for SMGs (with typical redshifts $\zs \sim 2-3$ and
deflectors at $\zl\sim 0.6$) than for SLACS lenses (for which $\zs\sim
0.7$ and $\zl\sim 0.2$).  Consequently, the deflector in \lnameII~and
similar $z\sim 0.6$ systems will typically have twice as large an
Einstein radius as lower redshift SLACS lenses, regardless of
differences in the deflector mass. This large radius implies that
these lenses will presumably be more dark matter
dominated. Extrapolating our best fit density profile inward to the
effective radius yields a projected dark matter fraction $f_{\rm
  dm}(<R_{\rm eff}) = 0.47 \mypm{0.19}{0.26}$, in agreement with SLACS
findings.

\section{Conclusions}\label{sec:conc}
We have calculated a detailed lens model of the newly found
gravitational lens \lname. Taking advantage of the best available
spatial resolution images in the \kband-band~using the Laser Guide
Star Adaptive Optics system at the Keck~II telescope, in the $i$-band
with the Subaru telescope and using Plateau de Bure Interferometer to
probe the CO($J=5\rightarrow4$) transition line at 576 GHz, we were
able to infer the mass distribution in the inner $\sim 30\kpc$ of the
deflecting structure, which turns out to be a massive galaxy,
presumably at the center of a galaxy group size halo of characteristic
velocity dispersion $\sigma_v=[483\pm 16] \,\kms$.

The redshift distribution of lensed SMGs will naturally select objects
in a favorable range for producing large Einstein radii, compared to
local lenses. This provides a great opportunity to probe the total
density profile and the dark matter content of intermediate redshift
halos with high accuracy. The lensing configuration of \lnameII~yields
a value of $R_{\rm Ein}=4\farcs05 \sim 27\kpc$ that corresponds to 6
times the effective radius of the central galaxy. At these scales we
observe a projected dark matter fraction of about 80\%.  The deflector
of \lnameII~probably stands in an interesting transition regime
between cluster and galaxy scale lenses, in which a joint lensing plus
dynamical analysis \citep{miralda95,sand04,gavazzi05,newman09} would
be very informative for the actual small-scale dark matter
distribution.

Much of the novelty of the large number of lensed SMGs to be uncovered
in ongoing and future submillimetric surveys resides in the
opportunity for studying faint heavily star-forming galaxies with good
resolution and signal-to-noise that would otherwise be unreachable.
In the particular case of \lnameII, the source turns out to be
magnified by a factor $\sim 10$, which allowed us to measure the
extent of the emitting gas in the CO($J=5\rightarrow4$) transition
line as well as young stars emitting in the NUV and visible bands. We
measured a half-light radius $R_{\rm eff,gas}= [1.13\pm0.53] \kpc$ and
$R_{\rm eff,*}= [2.0\pm0.1] \kpc$ using PdBI and Keck \kband~data,
respectively, in excellent agreement with the sample of SMGs studied
by \citet{swinbank10}.  There is some evidence that the peak of the CO
emission and the peak of the NUV/visible light could be offset by
$\sim 0\farcs3 \simeq 2.4 \kpc$, which could be explained if the
stellar light is obscured by dust associated with the cold gas. The
source reconstruction allows for a careful estimate of the intrinsic
source flux at many other wavelengths, thus allowing a more detailed
analysis of the gas and dust content of this peculiar SMG (C11, R11,
S11) and its gas content.

The large number of lensed SMGs like \lnameII~to be found in the
coming years with HerMES, H-ATLAS, the SPT and other surveys, will
allow detailed analyses of the central regions of massive galaxies
with exquisite accuracy over a broad range of deflector
redshifts. This is a good opportunity to constrain evolutionary trends
in their dynamical properties, including clues on the role of dry
vs.\ wet mergers or the role of the central AGN. At the same time, the
resolving power of gravitational lensing will allow detailed
investigations of the stellar, gaseous and dust content of massively
star forming SMGs at high redshift.


\acknowledgments

SPIRE has been developed by a consortium of institutes led by Cardiff
Univ. (UK) and including Univ. Lethbridge (Canada); NAOC (China); CEA,
LAM (France); IFSI, Univ. Padua (Italy); IAC (Spain); Stockholm
Observatory (Sweden); Imperial College London, RAL, UCL-MSSL, UKATC,
Univ. Sussex (UK); Caltech, JPL, NHSC, Univ. Colorado (USA). This
development has been supported by national funding agencies: CSA
(Canada); NAOC (China); CEA, CNES, CNRS (France); ASI (Italy); MCINN
(Spain); SNSB (Sweden); STFC (UK); and NASA (USA). The IRAM Plateau de
Bure Interferometer is supported by INSU/CNRS (France), MPG (Germany)
and IGN (Spain).




\end{document}